\documentclass[11pt,a4paper]{article}

\usepackage{authblk}

\usepackage[T1]{fontenc}
\usepackage[utf8]{inputenc}
\usepackage[ngerman,english]{babel}
\usepackage{graphicx}
\usepackage[square,numbers]{natbib}
\usepackage{tikz}
\usetikzlibrary{arrows, scopes, decorations.pathreplacing, decorations.markings, shapes, patterns, matrix, positioning} %external
\usepackage{pgfplots,pgfplotstable}

\title{Blending Entropy: A Term for Addressing Information Density in Mediated Reality}

\author[1]{Philipp Tiefenbacher\thanks{philipp.tiefenbacher@tum.de}}
\author[1]{Gerhard Rigoll\thanks{rigoll@tum.de}}
\affil[1]{Institute for Human-Machine Communication, Technical University of Munich}

\begin{document}

%%% This is the ``teaser'' command, which puts an figure, centered, below 
%%% the title and author information, and above the body of the content.

 %\teaser{
   %\includegraphics[height=1.5in]{images/sampleteaser}
   %\caption{Spring Training 2009, Peoria, AZ.}
 %}

\maketitle

\begin{abstract}
The virtuality continuum describes the degrees of positive virtuality under the umbrella term mixed reality. Besides adding virtual information within a mixed environment, diminished reality aims at reducing real world information. Mann defined the term mediated reality (MR), which also considered diminished reality, but without the possibility to describe different degrees of fusion between a mixed and a diminished reality. That is why this work defines the new term blending entropy that captures the relations between a mixed and a diminished reality. The blending entropy is based on the information density of the mediated reality and the actual area the user has to comprehend, which is named perceptual frustum. We describe the blending entropy's two-dimensional dependencies and detail important points in the blending entropy's space. 
\end{abstract}

 %\begin{CCSXML}
%<ccs2012>
%<concept>
%<concept_id>10003120.10003138.10003139</concept_id>
%<concept_desc>Human-centered computing~Ubiquitous and mobile computing theory, concepts and paradigms</concept_desc>
%<concept_significance>500</concept_significance>
%</concept>
%<concept>
%<concept_id>10003120.10003145.10011768</concept_id>
%<concept_desc>Human-centered computing~Visualization theory, concepts and paradigms</concept_desc>
%<concept_significance>500</concept_significance>
%</concept>
%</ccs2012>
%\end{CCSXML}
%
%\ccsdesc[500]{Human-centered computing~Ubiquitous and mobile computing theory, concepts and paradigms}
%\ccsdesc[500]{Human-centered computing~Visualization theory, concepts and paradigms}


% The next three commands are required, and insert the user-generated keywords, 
% The CCS concepts list, and the rights management text.
% Please make sure there is a blank line between each of these three commands.

%\keywordlist
%
%\conceptlist
%
%\printcopyright

\section{Definition of the Blending Entropy}
\begin{figure}[htb]
\vspace{-0.4cm}
\centering
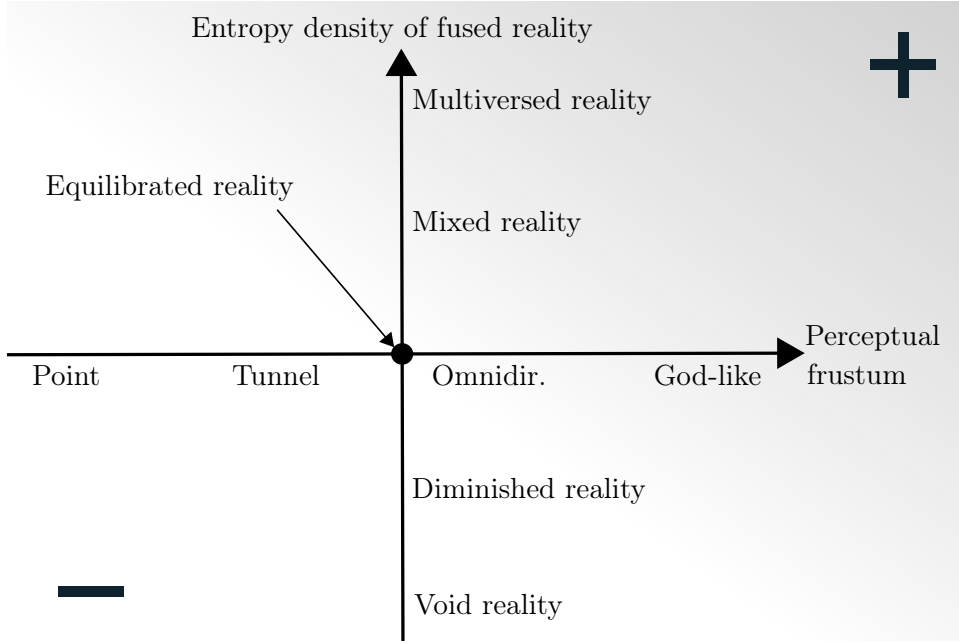 %mediatedreality_1.pdf_tex}
\caption{Illustration of the blending entropy for the perceived information density in a mediated reality (MR). The blending entropy is largest at the top right (+) and lowest at the bottom left (-).}
\label{int:rb}
\end{figure}
In information theory, entropy is the average information within a message. In a MR, entropy can be viewed as the total of any information perceived by the user. The blending entropy is characterized by the entropy density and the perceptual frustum as illustrated in Figure~\ref{int:rb}. 

\vspace{-0.145cm}
\subsection{Entropy Density}
\vspace{-0.11cm}
The information density or entropy density of an environment depends on the number and kind of physical objects and other visible information such as text, speech and haptic. The blending of the realities as defined in the virtuality continuum~\cite{Milgram.1995} fuses these entropy densities. Mann~\cite{Mann.1994} called it MR if this mixture can lead to a diminished reality. There is no clear concept yet that defines the stages between diminished and mixed reality. That is why the new term blending entropy is introduced.

The lower limit of entropy density is a reality with an entropy density of zero. The user that interacts in such a void reality would receive no information or knowledge. This void reality could be an environment that just consists of emptiness. Decreasing the entropy density in relation to the equilibrated reality leads to the diminished reality (DR), which aims at removing real-world objects unnoticeably, i.e, reducing the entropy density of the real world. 

The mixed reality comprises a larger entropy density than the equilibrated reality, since its goal is usually to add valuable information to the real- or virtual environment. The upper limit of the entropy density raises the uncertainty to a maximum, i.e., the environment only comprises independent and dense information. Such an environment can be accomplished by combining the real with (multiple) virtual realities into a multiversed reality. This definition of a multiversed reality motivates the importance of the term blending entropy. A multiversed reality would overwhelm the user's mind, i.e., the user would not be able to interact and reasonably process the information regardless of the size of the perceptual frustum. This is also called \textit{information overload} and describes the overmuch processing of information~\cite{Speier.1999}. % a state of overmuch information has to be processed at once.
This issue can be addressed via the blending entropy. For instance, minimizing the blending entropy circumvents an \textit{information overload}, which can be achieved by reducing the entropy density or the perceptual frustum. %  unable to process pushes the   In the depicted case, this system would be simply the human.

\vspace{-0.145cm}
\subsection{Perceptual Frustum} 
\vspace{-0.11cm}
The perceptual frustum is similar to the field of regard, but the perceptual frustum is dynamic. It depends on the situation and the spatial distribution of the necessary information. If the user has to be aware of information distributed in a vast environment for a certain task or purpose, the perceptual frustum is large. When compressing the required information into a small space, the perceptual frustum is tiny. 

The largest perceptual frustum is dubbed ``god-like''. It expresses a (theoretical) perceptual frustum that requires the user to concurrently access the total information of the world. In the work of Stafford et al.~\cite{Stafford.2006}, a ``god-like'' user tagged navigation clues on a tabletop for a user that received them in the real world via augmented reality (AR). This work is close to but still different from a ``god-like'' perceptual frustum, since the ``god-like'' user just focused on a single person and it was not necessary to conceive the whole reality. 

A perceptual frustum smaller than the ``god-like'' one is an omnidirectional perceptual frustum. The difference to ``god-like'' is that only the immediate environment of the virtual avatar or the person has to be comprehended and it is not necessary to consider multiple places of the world. 
Reducing the size of the perceptual frustum even further results in a tunnel-like perceptual frustum, which presents the necessary information right in front of the user. The ParaFrustum proposed by Sukan et al.~\cite{Sukan.2014} is a practicable example of a tunnel-like perceptual frustum. The ParaFrustum did not add further expert knowledge but it guided the user's view to the important parts of the current manual step.  

The smallest (theoretical) perceptual frustum is a point-like spot. A scenario of a user that has to press a lone button depending on the augmentation would comply to a point-like perceptual frustum. A study by Marner et al.~\cite{Marner.2013} is close to such a perceptual frustum. Here, the users pressed buttons sequentially according to spatial AR cues. 

%augmented button %not necessary to be aware of the whole world.
% or the area of interest is very smalnot the whole modelled environment may be perceived by the user, since the user may concentrate on a specific small part of the viewing frustum. The perceptual frustum defined the space that is actually of inter  which is perceived by the user. 
\subsection{Blending Entropy} 
\vspace{-0.11cm}
Finally, the blending entropy is the integration of the entropy density along the perceptual frustum. It holds the information of the joint realities that have to be perceived by the user. An increased or decreased blending entropy states that more or less information has to be processed by the user, respectively. The dotted line in Figure~\ref{int:rb} illustrates the direction of raising blending entropy. In consequence, the largest blending entropy exists for a multiversed reality with a ``god-like'' perceptual frustum.

\vspace{-0.145cm}
\subsection{Equilibrated Reality} 
\vspace{-0.11cm}
It is important to note that the origin of the blen\-ding entropy is defined by the entropies of the extrema of the virtuality continuum, i.e., the real or virtual environments.  A blending entropy of zero is called equilibrated reality. Here, the entropy density and the perceptual frustum of either the purely real or the purely virtual environments is identical to the density and the frustum of the equilibrated reality. Therefore, the equilibrated reality shares the same perceptual frustum and information density than the unmixed environments but it still consists of a modulated environment. Thus, the equilibrated reality can simply be described as a modulated reality that shares the identical information complexity as the purely virtual or real environments.
%mixes the contents of all available realities, e.g., just virtual content in the vr}. % The result defines the actual reality entropy for the user.

\vspace{-0.185cm}
\section{Examples}
\vspace{-0.11cm}
One of the goals of this work is to link mixed and diminished reality within a single representation. Mixed reality typically describes techniques and applications that advance the reality. The extension of the realities occurs for both extrema of the virtuality continuum. On the one hand, physical objects enhance the virtual reality, which is known as augmented virtuality~\cite{Milgram.1995}. On the other hand, virtual objects are integrated into the real world, i.e., AR. In every case, additional knowledge or material complements the environment, which introduces a higher entropy density according to the new terms of this work. 

Then reducing the entropy density connects the DR with the MR. An example for a DR application is the work of Korkalo et al.~\cite{Korkalo.2010}. They concealed AR markers via DR from the real environment, i.e., the goal was to lower the entropy density. Diminishing contents in the environment, while keeping the same perceptual frustum realizes a lower blending entropy.  
A term that describes all degrees of blending entropy already exists and is named mediated reality~\cite{Mann.1994}. The mediated reality consists of positive (MR) and negative (DR) entropy densities. The shortcoming of the original definition of the mediated reality is that there are no levels between mixed and diminished reality that describes the underlying relation. The definition of the entropy density allows such a classification. 

%Figure~\ref{int:rb} depicts the blending entropy as dashed axis. It can be seen that a large perceptual frustum with a high entropy density of the mediated environment yield maximum blending entropy.
% in an unchanged viewing frustum. This lea

The size of the blending entropy depends on the user as well as on the task. For example, the real environment of an industrial worker may already comprise a lot of information such as a console with different gauges, buttons and switches, i.e., the entropy density of the non-modulated environment is rather large. The worker, however, is accustomed to this environment and the specific tasks therein. This knowledge of the worker about the scene may produce a tunnel view, since the worker exactly knows where to look at, which minimizes complexity and creates an acceptable entropy for the worker. Consequently, the goal of a mediated reality can be the increase of the blending entropy and certain expert knowledge can be presented to the worker.

On the other hand, an untrained user that has not worked in such an environment before may be near an \textit{information overload}. Because of the definition of the blending entropy, also a variety of options are available for this user. The mediated reality application could aim at reducing the blending entropy or, at least, produce a balanced blending entropy, i.e., stay at the equilibrated state. 
This can be achieved via an AR-guided reduction of the perceptual frustum \cite{Sukan.2014}, which narrows the user's perception (tunnel vision). Moreover, the entropy density can be minimized via DR, i.e., the application identifies the actual relevant parts and diminishes the unimportant parts. Diminishing parts that are close to the border of the perceptual frustum may also decrease the perceptual frustum, which would generate an even smaller blending entropy.  

\vspace{-0.145cm}
\section{Conclusion}
\vspace{-0.11cm}
Most MR applications aim at modifying the blending entropy. For instance, some MR applications guide the user (lowering perceptual frustum) or present special information (higher entropy density), which can now be expressed via the blending entropy. Furthermore, the introduction of entropy density allowed to define a relationship between diminished and mixed reality. 

%\printbibliography
\bibliography{./2016_VRST_blending}

\begin{thebibliography}{7}
\providecommand{\natexlab}[1]{#1}
\providecommand{\url}[1]{\texttt{#1}}
\expandafter\ifx\csname urlstyle\endcsname\relax
  \providecommand{\doi}[1]{doi: #1}\else
  \providecommand{\doi}{doi: \begingroup \urlstyle{rm}\Url}\fi

\bibitem[Korkalo et~al.(2010)Korkalo, Aittala, and Siltanen]{Korkalo.2010}
O.~Korkalo, M.~Aittala, and S.~Siltanen.
\newblock Light-weight marker hiding for augmented reality.
\newblock In \emph{Proceedings of the International Symposium on Mixed and
  Augmented Reality}, pages 247--248. IEEE, 2010.

\bibitem[Mann(1994)]{Mann.1994}
S.~Mann.
\newblock Mediated reality.
\newblock Techreport 260, M.I.T. Media Lab Perceptual Computing Section, 1994.

\bibitem[Marner et~al.(2013)Marner, Irlitti, and Thomas]{Marner.2013}
M.~R. Marner, A.~Irlitti, and B.~H. Thomas.
\newblock Improving procedural task performance with augmented reality
  annotations.
\newblock In \emph{Proceedings of the International Symposium on Mixed and
  Augmented Reality}, pages 39--48. IEEE, 2013.

\bibitem[Milgram et~al.(1995)Milgram, Takemura, Utsumi, and
  Kishino]{Milgram.1995}
P.~Milgram, H.~Takemura, A.~Utsumi, and F.~Kishino.
\newblock {Augmented reality: A} class of displays on the reality-virtuality
  continuum.
\newblock In \emph{Proceedings of the Conference on Photonics for Industrial
  Applications}, pages 282--292. SPIE, 1995.

\bibitem[Speier et~al.(1999)Speier, Valacich, and Vessey]{Speier.1999}
C.~Speier, J.~S. Valacich, and I.~Vessey.
\newblock The influence of task interruption on individual decision making: An
  information overload perspective.
\newblock \emph{Wiley Online Library Decision Sciences}, 30\penalty0
  (2):\penalty0 337--360, 1999.

\bibitem[Stafford et~al.(2006)Stafford, Piekarski, and Thomas]{Stafford.2006}
A.~Stafford, W.~Piekarski, and B.~H. Thomas.
\newblock Implementation of god-like interaction techniques for supporting
  collaboration between outdoor ar and indoor tabletop users.
\newblock In \emph{Proceedings of the International Symposium on Mixed and
  Augmented Reality}, pages 165--172. IEEE, 2006.
\newblock \doi{10.1109/ISMAR.2006.297809}.

\bibitem[Sukan et~al.(2014)Sukan, Elvezio, Oda, Feiner, and
  Tversky]{Sukan.2014}
M.~Sukan, C.~Elvezio, O.~Oda, S.~Feiner, and B.~Tversky.
\newblock {ParaFrustum}: {V}isualization techniques for guiding a user to a
  constrained set of viewing positions and orientations.
\newblock In \emph{Proceedings of the Symposium on User Interface Software and
  Technology}, pages 331--340. ACM, 2014.
\newblock ISBN 978-1-4503-3069-5.

\end{thebibliography}

\end{document}